\documentclass[letterpaper, 10 pt, conference]{ieeeconf}  

\IEEEoverridecommandlockouts                              

\usepackage{tabularx}
\usepackage{makecell}
\usepackage{graphicx}
\usepackage{subfig}
\usepackage[dvipsnames]{xcolor}
\usepackage{hyperref}
\usepackage{caption}
\usepackage[font=footnotesize]{caption}
\usepackage{amsmath}
\usepackage [autostyle, english = american]{csquotes}
\MakeOuterQuote{"}

\title{\LARGE \bf
Investigating External Interaction Modality and Design Between Automated Vehicles and Pedestrians at Crossings}

\author{Sue Bai$^{1}$, Dakota Drake Legge$^{1}$, Ashley Young$^{1}$, Shan Bao$^{1,2}$, and Feng Zhou$^{1}$

\thanks{$^{1}$Sue Bai, Dakota Drake Legge, Ashley Young, Shan Bao, and Feng Zhou are with Department of Industrial and Manufacturing, Systems Engineering, University of Michigan, Dearborn, 4901 Evergreen Rd, Dearborn, MI 48128, USA
        {\tt\small \{baix, dlegge, ashleyly, shanbao, fezhou\}@umich.edu}}%
\thanks{$^{2}$Shan Bao is also with University of Michigan Transportation Research Institute, Ann Arbor, 2901 Baxter Rd, Ann Arbor, MI 48109, USA
       }%
}
\begin{document}

\maketitle
\thispagestyle{empty}
\pagestyle{empty}

\begin{abstract}

In this study, we investigated the effectiveness and user acceptance of three external interaction modalities (i.e., visual, auditory, and visual+auditory) in promoting communications between automated vehicle systems (AVS) and pedestrians at a crosswalk through a large number of combined designs. For this purpose, an online survey was designed and distributed to 68 participants. All participants reported their overall preferences for safety, comfort, trust, ease of understanding, usability, and acceptance towards the systems. Results showed that the visual+auditory interaction modality was the mostly preferred, followed by the visual interaction modality and then the auditory one. We also tested different visual and auditory interaction methods, and found that “Pedestrian silhouette on the front of the vehicle” was the best preferred option while middle-aged participants liked “Chime” much better than young participants though it was overall better preferred than others. Finally, communication between the AVS and pedestrians’ phones was not well received due to privacy concerns. These results provided important interface design recommendations in identifying better combination of visual and auditory designs and therefore improving AVS communicating their intention with pedestrians.

\end{abstract}

\section{INTRODUCTION}

Automated driving is expected to bring many societal benefits, such as decrease in vehicle crashes and increase in mobility \cite{ayoub2020modeling}. The development of automated vehicle systems (AVS) is progressing rapidly with options of different levels of automation. The SAE standards \cite{sae2018taxonomy} help AVS users, operators, and manufacturers to reach a common expectation of the vehicle’s capability and behavior from the vehicle control perspective. However, on-road traffic does not just consist of AVS. Other road users, such as pedestrians, cyclists, and motorcyclists also need to share the road system and therefore interact with AVS, especially in situations and maneuvers that need confirmation and/or negotiation, such as pedestrian crossing, four-way stop sign priority, and freeway merge situations. Pedestrians are among the most vulnerable road users \cite{ayoub2019manual}. Therefore, the safety of pedestrians is one of the most important safety areas to address for AVS design.

Pedestrians typically interact with human drivers to exchange their intention through eye-contact, gestures, verbal, approaching speed, and relative distance \cite{onishi2018survey,rasouli2019autonomous}. Therefore, researchers explored different ways by making use of these channels for the AVS to communicate with pedestrians. Förster et al. \cite{forster2011anthropomorphic} found that the participants in their study preferred a combination of the anthropomorphic design exterior and a list of functions, where eye-contact and mouth were important for humans to understand the robot. Jaguar’s Land Rover used a pair of eye-looking displays to mimic direct eye contact with the pedestrians \cite{rouchitsas2019external}, which also proved to be useful to establish mutual trust between the AVS and pedestrians \cite{de2019perceived}.

Another common method is to directly use verbal information (e.g., “Walk”, “Don’t Walk”). For example, de Clercq et al. \cite{de2019external} found that verbal information delivered by a textual display was regarded as the least ambiguous among others. Mahadevan et al. \cite{mahadevan2018communicating} used a verbal message (“I see you”) with a speaker to communicate AVS' awareness and three LEDs (yellow, red, and green) to show its intent. The interface made use of both auditory and visual modalities and were found to be most effective among the four tested concepts. Wu et al. \cite{wu2014cars} leveraged dedicated short-range communications developed by Honda and Qualcomm through smartphones to preempt a possible collision between a pedestrian and an approaching vehicle. Both visual (bright yellow color with full screen) and auditory signals (car horns and high-pitched beeps) were used to deliver alerts to pedestrians.  

Several researchers \cite{onishi2018survey,rasouli2019autonomous} conducted comprehensive reviews of existing work on the interaction methods sorted by technology, location, content, modality, and experimental tasks. Based on their results, the majority of the previous studies mainly focused on exploring visual modalities with a limited number of auditory warnings. In addition, there is no common protocols or a set of characteristics for standardizing designs across the industry and the academia. Furthermore, pedestrians’ reaction to AVS movements were also varied by regions between metropolitan areas and small coastal cities in Mexico \cite{currano2018vamos}. If various AVS have non-consistent methods to communicate with all pedestrians, it can cause confusion to all the road users \cite{deb2018pedestrians}. Therefore, an easy-to-understand and standardized approach for the AVS to interact with pedestrians would be necessary to improve road safety and efficiency by examining a large number of designs \cite{ayoub2019manual}.

In this study, we aimed to test the effectiveness of different combinations of three modalities, including visual, auditory, and visual+auditory through video presentations and to identify the optimal design for AV-pedestrian interaction. More specifically, this study was designed to identify the best visual and auditory warning displays by examining six visual designs (cross-bar, pedestrian silhouette, text display in front and on the roof, face emojis, and no display), and five types of auditory alerts (beep, chime, honk, human voice, and no audio). We also investigated if there were any age-related differences in user acceptance.

\section{METHOD}

\subsection{Participants}

To achieve the objectives, we designed and conducted a survey study using Qualtrics (Provo, UT).  A total number of 68 participants (33 females and 35 males) from the United States participated in this survey study. The age distribution of the participants was as follows: 27\% were between 16-23, 34\% were between 24-39, 27\% were between 40-54, and 12\% were 55 or older. All the participants had a valid US driver’s license and they were not compensated upon their completion of the survey. One participant’s data was removed due to incomplete data. Therefore, data from the remaining 67 participants were used in the final analysis. Our study was approved by the Institutional Review Board at the University of Michigan. 

\begin{figure} [thpb]
\centering
\includegraphics[width=1\linewidth]{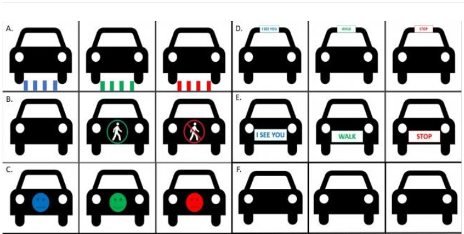}
\caption{AVS-pedestrian visual interaction inquiry options.}
\label{fig:1}
\end{figure}

\subsection{Experiment design}

\textbf{Independent variables.} In this study, we conducted a six by five within-subjects experiment. The first within-subject variable was visual interaction designs (see Fig. 1) and the second within-subject variable was auditory alert types (chime, beep, voice, honk, and no-audio). This design allowed us to examine participants' preferences on AVS-pedestrian interaction modalities, i.e., visual, auditory, and visual+auditory as well as design combinations. Note there was a control condition for the visual and auditory designs, respectively, i.e., no visual or auditory information, which was used to test the auditory or visual modality alone. Furthermore, we also tested whether they preferred to send alerts to their smart phones or not. The order of independent variables were presented randomly to the participants. Besides, we considered age as another variable in the analysis process by dividing the participants into two age groups, young ($\leq$ 39 years old, i.e., millennial or younger, sample size = 41) and middle-aged group ($\geq$ 40 years old, sample size = 36). 

\textbf{Dependent variables.} Based on the interaction designs presented to the participants, we ranked their overall preferences and measured their perception on safety, comfort, trust, ease to understand, usability, and acceptance using a 7-point Likert scale (1 = strongly agree, 2 = agree, 3 = somewhat agree, 4 = neither agree nor disagree, 5 = somewhat disagree, 6 = disagree, 7 = strongly disagree). Participants were also required to pick the most preferred one among six visual interaction designs, five auditory alerts, and options to sending alerts on smart phones or not. In addition, participants were asked to state the reason for their preference selections as an open-ended question at the end of the survey.

\textbf{Survey design.} The survey had four sections. The first section consisted of a consent form describing the introduction and the purpose of the study; the second was designed to collect participants' demographic information; the third section provided all the combinations of interaction designs in videos in a random order, with the same pedestrian crossing scene. The participants were asked to watch the video and answer the questions followed, which were used to measure the dependent variables introduced above. The fourth section required the participants to pick one of their most preferred one among six visual interaction designs (see Fig. 1), five auditory alerts (chime, beep, voice, honk, and no-audio), and the option of sending alerts on smart phones or not, respectively. The estimated time to finish the survey was about 45 minutes.

\subsection{Procedure}
First, each participant went through the four sections of the survey as described above. Then, the participants were asked to state the reasons on their selection optionally. Since there were six visual interaction designs and five auditory alerts, the participants were requested to evaluate a total number of 30 combinations and give the ranks of the six visual and five auditory designs, respectively. 

\subsection{AVS-Pedestrian Interaction Scenarios in Video}
As an example, Fig. 2 shows the screen shots of the animation videos of the pedestrian crossing scenarios with one combinations of one visual (i.e., cross-bar projection on the ground in front of AV) and audio (i.e., chime) combination. The videos can be found by the following links:

\begin{itemize}
\item Visual: \url{https://youtu.be/cZQ7OMhkWwE}
\item Auditory: \url{https://youtu.be/8oAJuucjRfs}
\item Visual+auditory: \url{https://youtu.be/ZFwVM1ezhMU}
\end{itemize}

In the visual condition, the video first 1) shows a pedestrian standing at the side of a crossing point and the AVS is stopping as shown in Fig. 2(a),  then 2) the AVS projects a green crossing-bar (“allowed to cross”, see Fig. 2(b)) on the ground with a pedestrian walking, 3) the crossing-bar turns yellow (“caution to cross”, see Fig. 2(c)), and finally 4) to red (“not allowed to cross”, see Fig. 2(d)) without a pedestrian present. In the auditory condition, the AVS gives 1) three sets of fast-paced chimes with one tone (“getting ready to cross”), 2) one continuous set of slow-paced chimes with two tones (“allowed to cross”), 3) one set of fast-paced chimes with two tones (“caution to cross”), and 4) one set of fast-paced chime with one tone (“not allowed to cross”). In the visual+auditory condition, the three-colored crossing-bars in the visual condition and four sound effects in the auditory condition are overlaid in order. When no visual or no auditory alert was combined with another auditory or visual design shown to the participants, they were considered auditory only and visual only. 

\begin{figure} [bt!]
\centering
\includegraphics[width=1\linewidth]{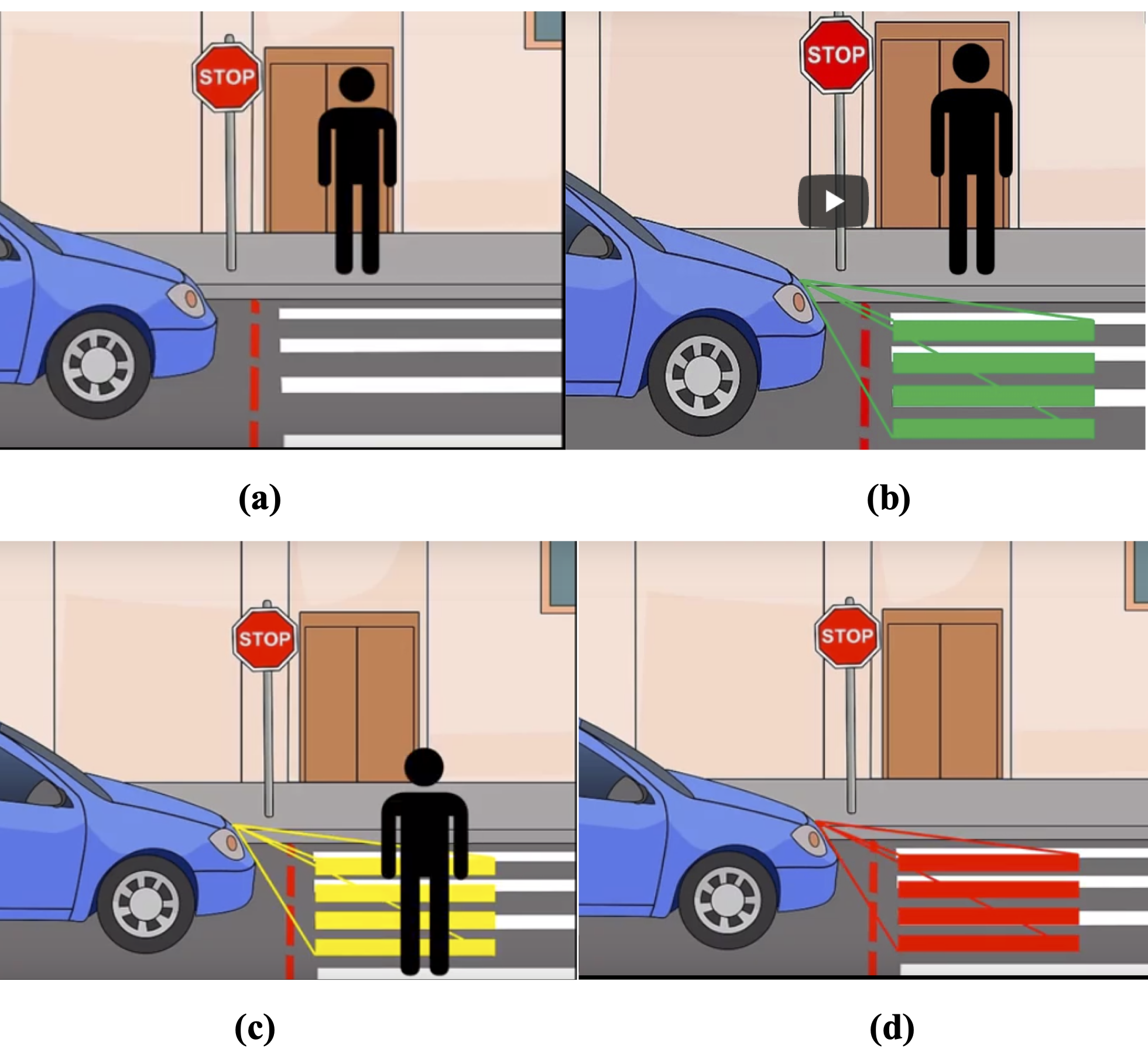}
\caption{Video screen sample of the visual condition.}
\label{fig:2}
\end{figure}

\subsection{Data Analysis}
Due to the fact that the sample sizes of the young group and the middle-aged group were not the same and the Friedman test was not able to analyze multiple factors or interactions, we used the Aligned Rank Transform (ART) for nonparametric factorial data analysis \cite{wobbrock2011aligned}. ART first aligned data before applying averaged ranks and then the common ANOVA procedures were used. The RStudio software and the ARTool package was used to conduct the statistical analysis below. Note in this paper, the results mainly focused on the modality results across different combinations of designs between visual and auditory modalities. 

\section{RESULTS}
\subsection{Comparison Among Three Modalities}

First, two-way ANOVA after ART was used to test overall preference among three modalities and we found a significant main effect for modality $F(2,130)=38.52,p=0.000$, but no effects for age or interaction effect as shown in Fig. 3(a). Then we conducted post-hoc analysis with Turkey HSD correction and found the visual+auditory condition was ranked significantly higher than auditory $(p=0.000)$ and visual $(p=0.001)$ conditions and the visual condition was ranked significantly higher than the auditory condition $(p=0.006)$.

\begin{figure} [bt!]
\centering
\includegraphics[width=1\linewidth]{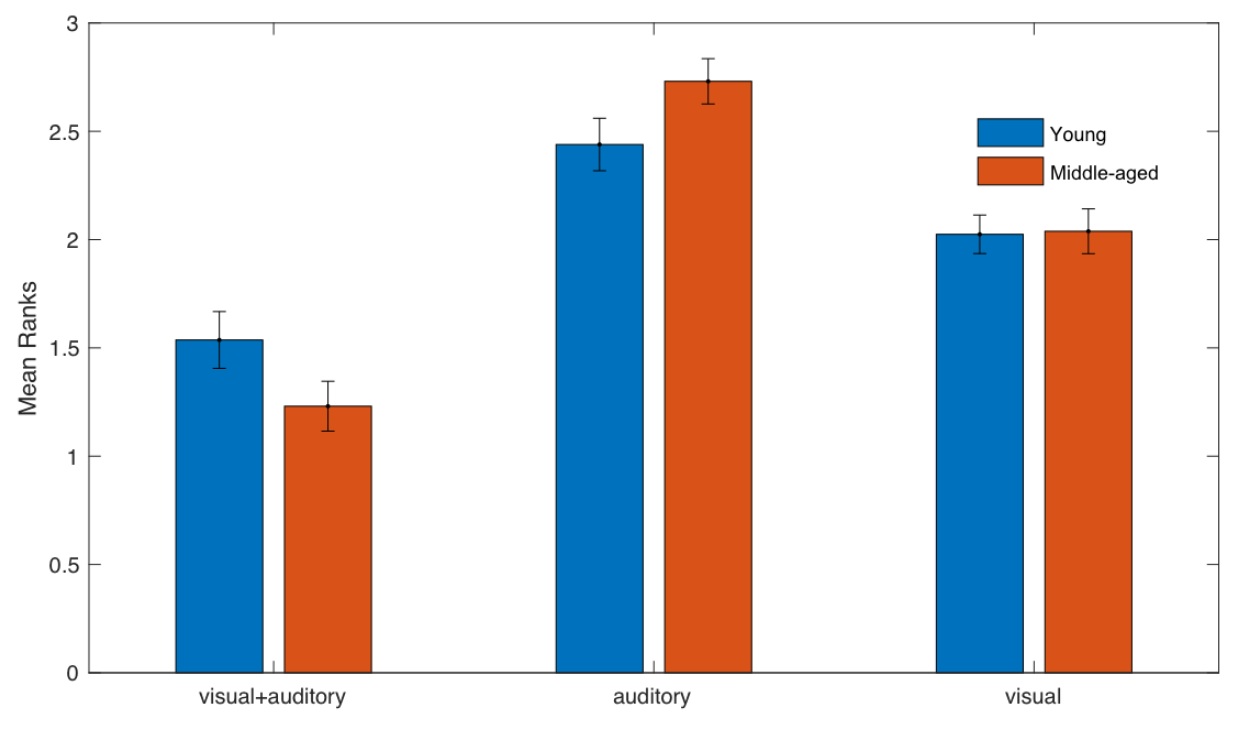}
\caption{Preference mean ranks with standard errors of three interaction scenarios and age, where “1” ranks the best and “3” ranks the worst.}
\label{fig:3}
\end{figure}

\begin{figure} [bt!]
\centering
\includegraphics[width=1\linewidth]{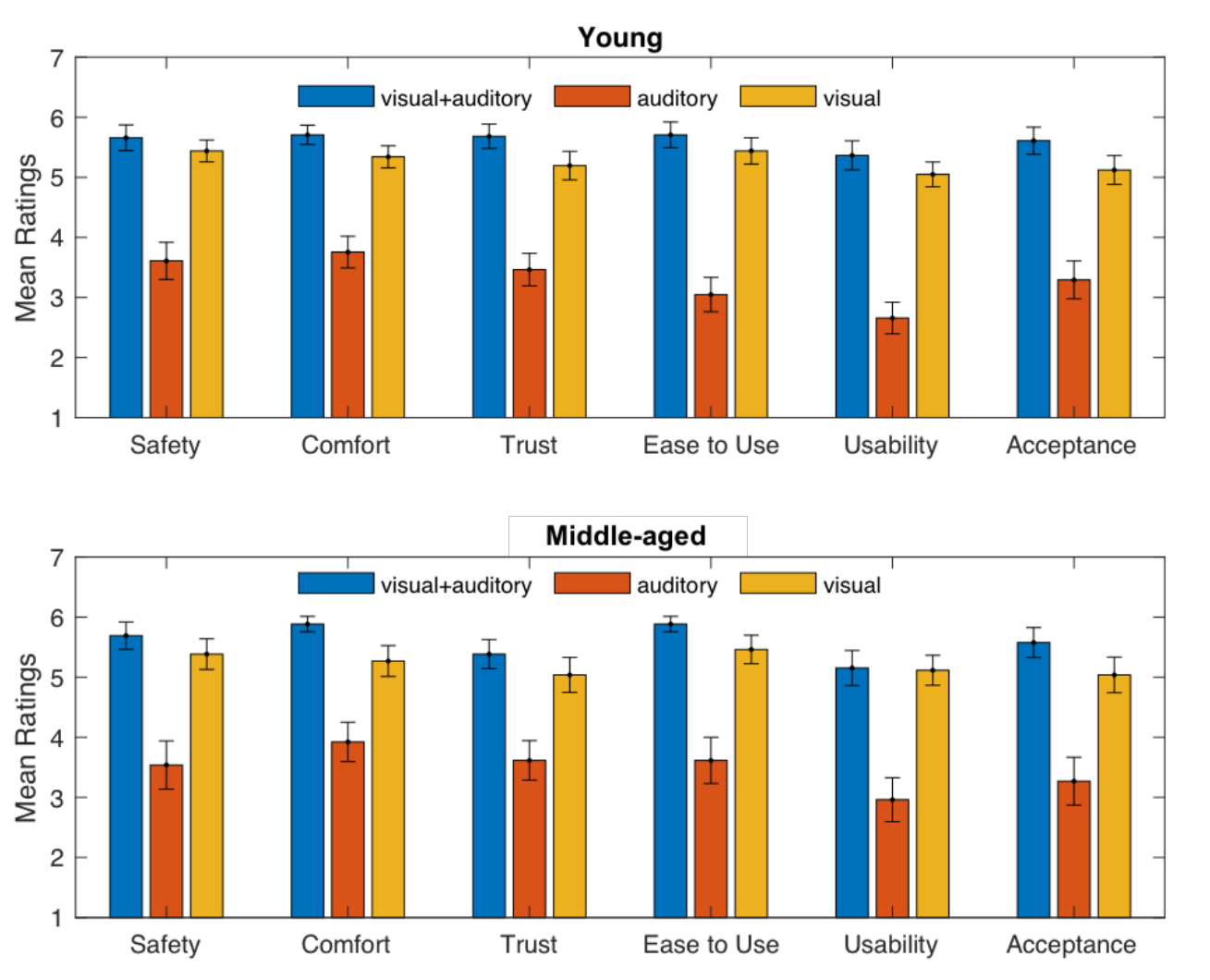}
\caption{Mean ratings with standard errors of individual dependent variables of three interaction scenarios, where “7” indicates strongly agree and “1” strongly disagree.}
\label{fig:4}
\end{figure}

Second, the participants were also asked to rate other measures, including safety, comfort, trust, ease to understand, usability, and acceptance on a 7-point Likert scale. The results were shown in Fig. 4. Two-way ANOVA after ART was used for each of the function measures assessment. For all the function measures, a significant main effect was found (Safety:  $F(2,130)=42.25,p=0.000$; Comfort: $F(2,130)=38.09,p=0.000$; Trust: $F(2,130)=51.81,p=0.000$; Ease to Understand: $F(2,130)=56.53,p=0.000$; Usability: $F(2,130)=70.45,p=0.000$; Acceptance: $F(2,130)=39.22,p=0.000$). No main effects were found for age nor interaction effects. Post-hoc analysis with Turkey HSD correction showed that both visual+auditory and visual were significantly rated better than auditory (all $p=0.000$). There was only one marginal difference between visual+auditory and visual for Comfort ($p=0.084$). 

\subsection{Preferences of Visual Interaction Designs} 

The participants were asked to pick (indicated by “1”, otherwise “0”) the most preferred one among six visual designs from Fig. 1. We used the two-way ANOVA after ART and found that there were a significant main effect for designs, $F(5,325)=12.73,p=0.000$ and no main effect for age or interaction effect was found as shown in Fig. 5. Post-hoc analysis with Turkey HSD correction found that Option B (Pedestrian silhouette on the front of AV) was ranked significantly better (all p=0.000) than all other options, except option A (Cross-bar projection on the ground in front of AV). Option A ranked significantly better than option C (Face emoji on front of AV, $p=0.020$), option D (Text display on the roof of AV, $p=0.002$), and option F (no visual indication). Option E (Text display on front of AV) ranked significantly better than option F.
\begin{figure} [bt!]
\centering
\includegraphics[width=1\linewidth]{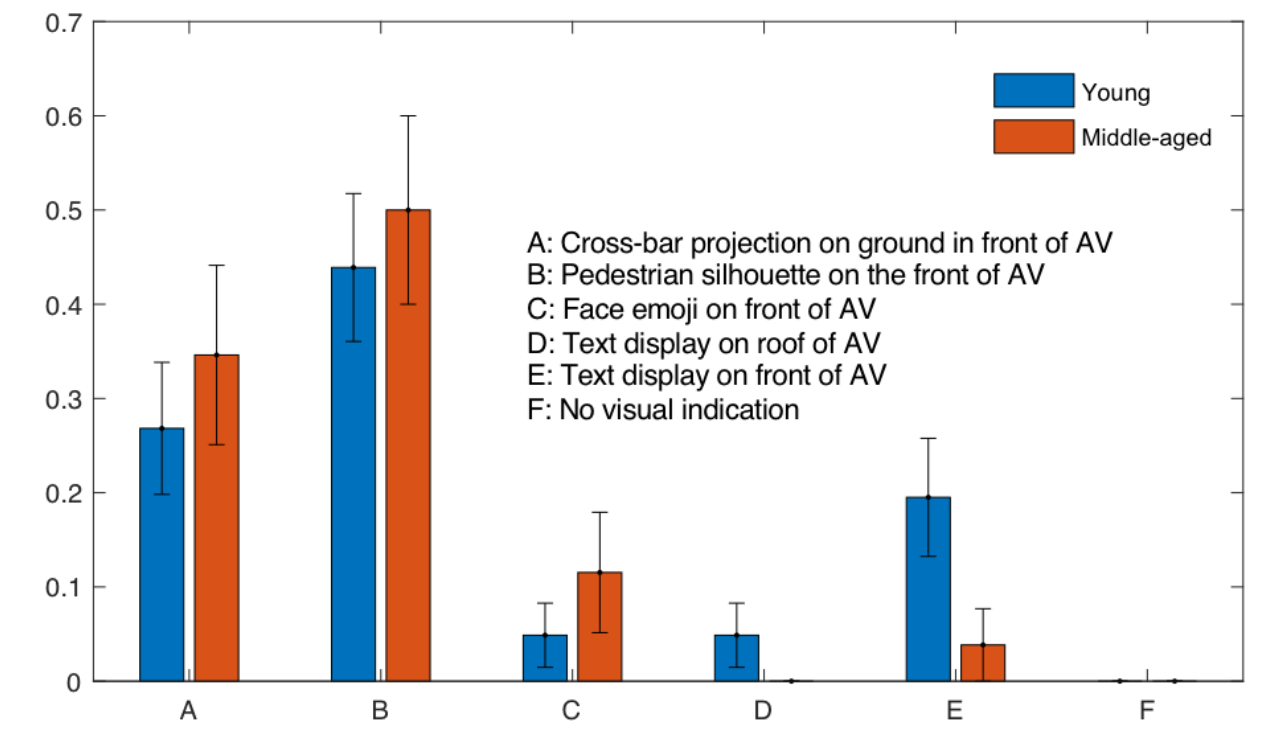}
\caption{Preference mean ranks with standard errors of six visual designs, where “1” ranks the best and “0” ranks the worst.}
\label{fig:5}
\end{figure}

\begin{figure} [bt!]
\centering
\includegraphics[width=1\linewidth]{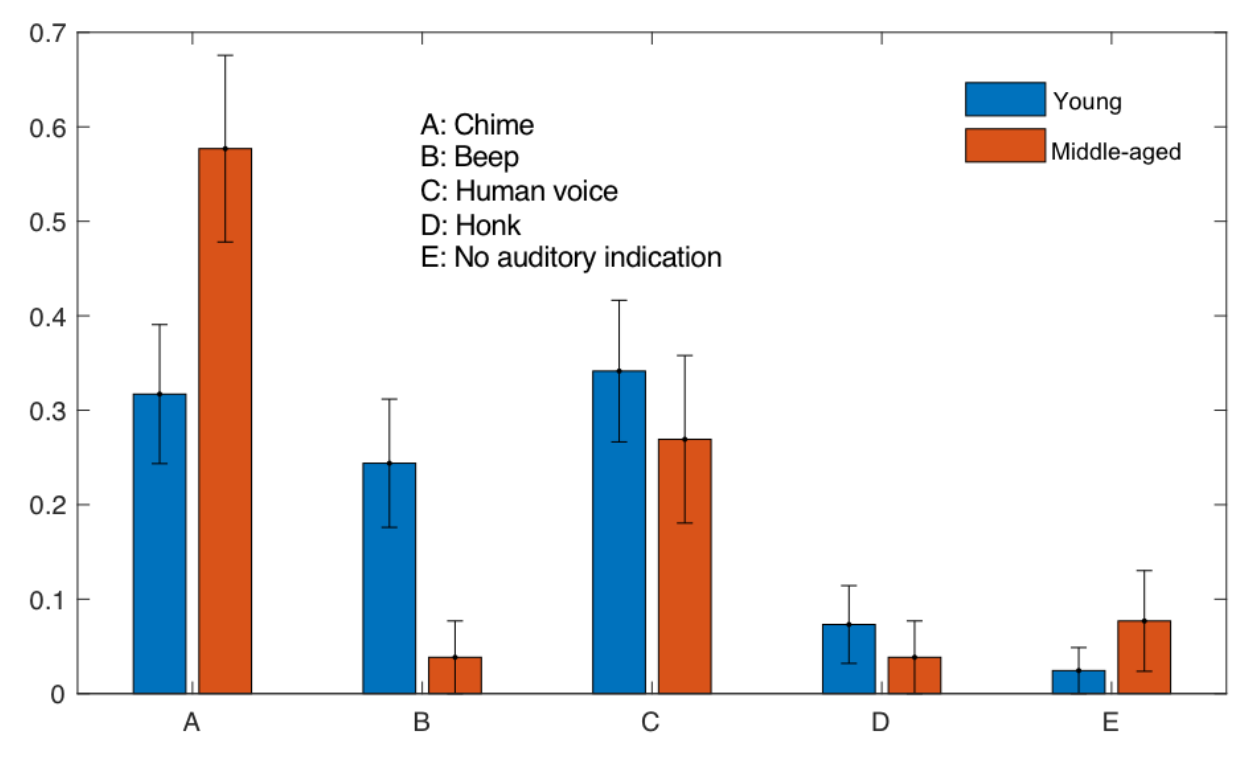}
\caption{Preference mean ranks with standard errors of five auditory alerts, where “1” ranks the best and “0” ranks the worst.}
\label{fig:6}
\end{figure}

\begin{figure} [bt!]
\centering
\includegraphics[width=1\linewidth]{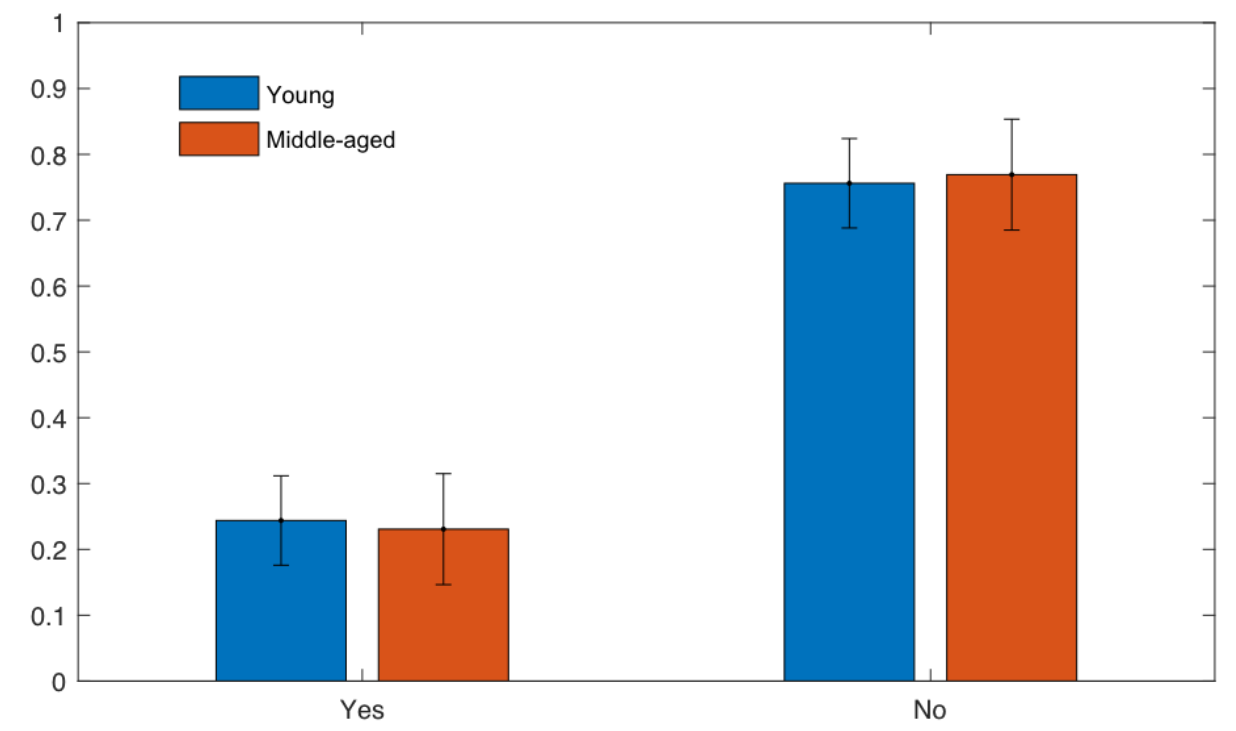}
\caption{Preference mean ranks with standard errors of smart phone alerts, where “1” ranks the best and “0” ranks the worst.}
\label{fig:7}
\end{figure}

\subsection{Preferences of Auditory Alerts }

Likewise, the participants were asked to pick (indicated by “1”, otherwise “0”) the most preferred one among five auditory alerts. We used the two-way mixed ANOVA after ART and found that there were a significant main effect for designs, $F(4,260)=15.82,p=0.000$ and a significant interaction effect $F(4,260)=3.70,p=0.006$ as shown in Fig. 6. Post-hoc analysis with Turkey HSD correction found that option A (Chime) was ranked marginally better than option B (Beep, $p=0.054$), and significantly better than option D (Honk, $p=0.008$) and option E (No auditory indication, $p<0.001$) and that option C ranked significantly better than option D ($p=0.038$) and option E ($p<0.001$). Furthermore, while young participants preferred Beep significantly better than middle-aged participants ($p=0.042$), middle-aged participants preferred Chime significantly better than young participants $p<0.001$.

\subsection{Preferences of Smart Phone Alerts}

Finally, the participants were asked to pick (indicated by “1”, otherwise “0”) whether they would prefer alerts on their smart phones. We used two-way mixed ANOVA after ART and found that participants preferred no smart phone alerts significantly better than smart phone alerts ($F(1,65)=14.83,p<0.001$) and that no effect was found for age as shown in Fig. 7.

\section{DISCUSSIONS}
\subsection{Auditory Cues} 
Our results showed that the visual+auditory modality was the most preferred one among the three tested modalities, followed by the visual modality and the auditory modality, which was consistent with previous studies \cite{mahadevan2018communicating}. In our study, auditory cues were not well received due to several reasons. 

First, the difference among different types of auditory alerts was not very clear. This might be because the participants had difficulty remembering or learning them with tones and speeds of auditory cues during the interaction process (P5: “\emph{Seems like there is room for error for people to forget which sound means what}”; P22: “\emph{I found this the most confusing. You would need to learn what the sounds mean}”; P64: “\emph{There are a lot of different tones and speeds of the chimes.  … it would take a while to get pedestrians to understand}”) or because there were no standardized auditory alerts among different automotive manufacturers (P7: “\emph{Different manufacturers can use different tones, making it hard to know what tone is what}”). 

Second, it might be difficult to hear the auditory alerts if the environment is noisy (P48: “\emph{Would be concerned about being able to hear the audible indication on a busy/noisy street}”; P50: “\emph{Sound may not be heard depending on ambient noise) or pedestrians had headphones/earphones on or were hearing-impaired}” (P53: “\emph{I worry about pedestrians who are deaf/hard of hearing, and those wearing headphones, ear bugs, winter hats, hoods, etc.}”).  

Among the five auditory alerts, chime and human voices tended to be the preferred ones overall (see Fig. 7). Chime was less aggressive or annoying and resembled some current cross walk signals better (P23: “\emph{Chiming seems less annoying}”; P44: “\emph{Again resembles some of the cross-walk types and would be the best to let people know}”) and human voices was least confusing and most informative though it was language dependent (P16: “\emph{can’t misinterpret}”; P28: “\emph{Think it would be most informative}”; P48: “\emph{Voice would be best/least ambiguous. Obviously would need to account for different languages, which could be a challenge}”). Honk was rated to be annoying although it was easy to understand its semantics (P5: “\emph{Honk for don't walk or at least some very obvious sound for do not go. (Note that this will be annoying in crowded cites)}”). 

We did notice a significant difference between two age groups in terms of chime and beep. Young participants (P5: “\emph{It's the most practical}”; P28: “\emph{beeps are small short and to the point}”; P38: “\emph{...beep should be enough}”) liked beep significantly better than middle-aged participants. Middle-aged participant (P13: “\emph{Not as alarming}”; P41: “\emph{Pleasant, calm, not annoying}”; P60: “\emph{More of a pleasant sound}”) liked chime significantly better than young participants. Such difference was mainly caused by the fact that middle-aged participants tended to prefer pleasant and calm alerts than annoying ones while young participants prefer beeps as they were practical.

\subsection{Visual Cues}
Compared to auditory alerts, participants preferred visual cues better, which might be due to the fact that they were more familiar with the semantics of the visual cues. However, one major concern for the visual interaction scenario was how to address those who were color blinded, visually impaired, or distracted (P64: “\emph{Much better than the audio only, but no good for blind people}”; P40: “\emph{only concern is everyone looks at their phone now, will they see and understand...also what about bad weather}”). In this aspect, designers might consider alternative modalities on top of visual cues while considering individual characteristics of pedestrians and constantly changing technology usage patterns. Another concern was that the participants were not sure that such visual projection with different colors on the ground would be clear in different weather conditions (P21: “\emph{Would it work in the rain/ glare of the sun just the same?}”; P66: “\emph{Sunny days may make the projection hard to see so a second indicator on the car would help}”). Therefore, such visual cues need to be fully tested in various conditions before it can be implemented in AVS. 

Although the visual+auditory condition was rated the best (P24: “\emph{Overall clear and easy to understand}”), the limitations in the visual (e.g., visibility in the sunny condition) or in the auditory (e.g., confusion between auditory cues) conditions could still be applied. Therefore, the designers should leverage complementary effects between visual and auditory cues in the design process.

Among different types of visual interaction designs (see Fig. 1 and Fig. 6), option B, i.e., Pedestrian silhouette on the front of the AV tended to be preferred better among all the tested conditions. This was mainly because the participants were familiar with the symbols and it was language independent (P6: “\emph{Clear and similar to existing symbols}”; P13: “\emph{Universal symbols}”; P29: “\emph{Clear, unambiguous, language independent}”) and the limitations of other designs (P7: “\emph{Projections are risky. Words are too much effort. Simple pictures are best, but smiley faces don't provide enough context}”). Compared to option B, option A (Cross-bar projection on the ground in front of AV) was also better received and sometimes was perceived better than option B. This was mainly because it was more natural to look at the road than at the vehicle (P23: “\emph{You can look at the road rather than the car}”; P28: “\emph{People normally look down or straight out in front them while walking}”). Although the text displays (option D or E) were probably least confused \cite{de2019external}, they were less universally recognized than familiar symbols, such as option A and option B (P59: “\emph{I think universal pictures are usually better than words}”). Generally, the participants tended to dislike the emoji face design (option C) mainly because it was not clear (P5: “\emph{…but smiley faces don't provide enough context}”; P48: “\emph{… I have no idea what the smiley face means}”). No significant difference for age or interaction effect was found which indicated that there was a general consensus among the two age groups.

\subsection{Alerts on smart phones} 

Alerts sent to pedestrians’ personal smart phones were not well received (see Fig. 4 (c)). This was mainly due to privacy concerns (P4: “\emph{I don’t want some strange vehicle connected to my phone}”; P16: “\emph{privacy concerns, and vehicle hacking}”), availability issues (P15: “\emph{What if my phone is put away and out of reach?}”) and distraction (P36: “\emph{Easy way to distract oneself}”). Although the vehicle-to-pedestrian wireless communication technology was designed as an ad hoc direct device-to-device communication without the needs to exchange phone numbers. This explanation was not explicitly stated in the survey. As a result, it led to the participants’ misunderstanding of the privacy intrusion and their negative feelings towards such a type of the communication.  Also, no significance between the two age groups or interaction effect was found. Thus, both of the two age groups had such privacy concerns.

\subsection{Limitations and future work} 
The web-based survey allowed us to reach a broader participant pool with both qualitative and quantitative data. Although we understood underlying reasons for their choices to a large extent using qualitative data, it still limited the value of the study due to the lack of in-person interaction with more questions and answers, which may lead to valuable input on additional preferences or concerns that the participants may express. In addition, we only surveyed participants within the U.S. with valid driver licenses using video interaction. Therefore, it should be cautious to generalize our results to other contexts. In the real world, not all pedestrians are capable of driving a vehicle and understanding the vehicle dynamics in general, and not all pedestrians can understand English. 

The age distribution was not very balanced, and not many old participants (age 65 and over) were recruited. Therefore, further research is needed to gain more insights into older and/or non-driver, non-English speaking pedestrian’s perception regarding intention communication with AVS.

Furthermore, it was well agreed that the design to facilitate communications between AVS and other road users should be standardized within the automotive industry in order to avoid confusion across various vehicle types. Therefore, it is important to reach some type of agreement throughout the automotive industry, government agencies, and the academia.

\section{CONCLUSIONS}

This study investigated pedestrian’s preference in the interface design for AVS to communicate its yield or proceeding intention to the pedestrian at a crosswalk. An online survey was created with three interaction modalities shown to the participants as an animation video: visual+auditory, auditory, and visual. The combination of audio and visual indication was the most preferred among the participants, followed by visual only and auditory only. In terms of visual interaction designs, the pedestrian silhouette on the front of the AV and the ground projection of the cross-walk bars were the most preferred. Soft chime and human voices were the preferred auditory designs. Due to the concern of privacy intrusion, the smartphone-based alerts were not preferred. Generally, there was not difference between the two age groups tested except the preferences to chime and beep. Our results gave guidelines about how to identify the best possible combinations of visual and auditory designs for the AV to communicate its intention to the pedestrian and showed important design implications for the future communication between AVS and pedestrians.

\addtolength{\textheight}{-12cm}   





\bibliography{root}

\end{document}